\documentclass[fleqn,10pt]{wlscirep}
\usepackage[utf8]{inputenc}
\usepackage[T1]{fontenc}
\title{Assessing the attraction of cities on venture capital from a scaling law perspective}
\author[1]{Ruiqi Li}
\author[1]{Lingyun Lu}
\author[1,*]{Weiwei Gu}
\author[1]{Shaodong Ma}
\author[2,*]{Gang Xu}
\author[3,*]{H. Eugene Stanley}

\affil[1]{UrbanNet Lab, College of Information Science and Technology, Beijing University of Chemical Technology, Beijing 100029, China}
\affil[2]{School of Remote Sensing and Information Engineering, Wuhan University, 129 Luoyu Road, Wuhan 430079, China}
\affil[3]{Center for Polymer Studies and Physics Department, Boston University, Boston, MA 02215, USA}
\affil[*]{corresponding. weiweigu@mail.buct.edu.cn, xugang@whu.edu.cn, hes@bu.edu}


\begin{abstract}
Cities are centers for the integration of capital and incubators of invention, and attracting venture capital (VC) is of great importance for cities to advance in innovative technology and business models towards a sustainable and prosperous future. Yet we still lack a quantitative understanding of the relationship between urban characteristics and VC activities. In this paper, we find a clear nonlinear scaling relationship between VC activities and the urban population of Chinese cities. 
In such nonlinear systems, the widely applied linear per capita indicators would be either biased to larger cities or smaller cities depends on whether it is superlinear or sublinear, while the residual of cities relative to the prediction of scaling law is a more objective and scale-invariant metric. 
Such a metric can distinguish the effects of local dynamics and scaled growth induced by the change of population size. The spatiotemporal evolution of such metrics on VC activities reveals three distinct groups of cities, two of which stand out with increasing and decreasing trends, respectively. And the taxonomy results together with spatial analysis also signify different development modes between large urban agglomeration regions. Besides, we notice the evolution of scaling exponents on VC activities are of much larger fluctuations than on socioeconomic output of cities, and a conceptual model that focuses on the growth dynamics of different sized cities can well explain it, which we assume would be general to other scenarios.
\end{abstract}
\begin{document}

\flushbottom
\maketitle
\thispagestyle{empty}

\section{Introduction}
Cities are centers for the integration of capital and incubators of invention, which have created more than 80\% wealth \cite{dobbs2011urban} and 90\% innovation \cite{rothwell2013patenting} worldwide, mainly due to agglomeration and knowledge spillover effect \cite{bettencourt2007invention,romer1986increasing,lucas1988mechanics,wang2020high,park2019global}, and played a crucial role in the development of science, technology, novel business models and culture. 
Partially due to the knowledge spillover effect and self-reinforcing process, cities with greater creation and higher concentration of both knowledge and capital would be more attractive to educated, highly skilled, entrepreneurial and creative individuals \cite{bettencourt2007invention,keuschnigg2019scaling,wang2020high,park2019global}, which has long been recognized as positive externalities generated from the increase of urban scale \cite{marshall2009principles,jacobs1969economy}. 

Venture capital (VC) is a form of private equity financing that is provided by venture capital firms/funds to startups, early-stage, and emerging companies that have been deemed to have high growth potential or which have demonstrated high growth. 
Venture capital investment activities have been largely an urban phenomenon since its institutionalization around 1980s \cite{hochberg2007whom}. Venture capital investment is usually regarded as the engine of inventions and wind indicator of emerging markets opened up by innovative technology or business model \cite{sun2019venture}, which is of high potential returns, as well as of high uncertainties, risks and failure rates. Attracting more venture capital investments in an industry is associated with significantly higher patenting rates and spurring of innovation \cite{kortum2001does,rothwell2013patenting}. 
A quantitative understanding of the relationship between urban characteristics and venture capital activities would be important for the development of cities towards a sustainable and prosperous future. 

From a physical perspective, the size of population would be the most important attribute as itself is a manifestation of the attraction and maintenance on financial and human capital \cite{bettencourt2007invention,bettencourt2007growth,bettencourt2013origins,li2017simple}. 
Various superlinear scaling laws have been discovered on socioeconomic related quantities, including GMP (Gross Metropolitan Product), income, the number of patents, severe crime and HIV/AIDS cases, in different countries over different periods of time \cite{bettencourt2007growth,bettencourt2013origins,gomez2016explaining,li2017simple,li2016geometric,xu2019paradoxical}. 
Yet currently, to our best knowledge, there has been quite few work focus on scaling analysis of venture capital activities -- e.g., whether the amount of venture capital investment will scale linearly or superlinearly with the urban population size. In addition, another compelling question is to ascertain what types of cities (or even more precisely, which features of urban systems) would attract more venture capital investments. An intuitive way might be looking at the total amount or the per capita value, yet such commonly used criteria assume the linearity of the system where the ensemble is just a linear sum of all its elements. However, we are well aware that cities are complex systems with nonlinear effects manifested as super- and sub-linear scaling relation with respect to the urban population. In such situations, it would be unfair and inherently biased to compare per capita value; for example, there has been a strong evidence showing that the total weight can be lifted by champion weightlifters is in a 2/3 sublinear relation with their body mass \cite{lietzke1956relation}, which means that if we compare per capita value, heavier weightlifters will be always underrated on a per capita basis \cite{west2017scale}. 
This suggest that we have to come up with a more objective measurement if there are nonlinear effects in the system. 

As the scaling law never appear by accident \cite{barenblatt2003scaling}, it can be regarded as a baseline or null model, similar to the criterion of judging who is the strongest weightlifter \cite{west2017scale}, where the residual relative to the prediction by the scaling law is a more objective and scale-invariant indicator \cite{bettencourt2010urban,zund2019growth}. Such idea has been developed into a standard measurement named Scale-Invariant Metropolitan Indicator (SIMI) \cite{zund2019growth,bettencourt2010urban,bettencourt2020urban}. From the evolution of SIMI of cities, we can further identify groups of cities with similar performance to reveal possible relationship between cities or even developing modes and dynamics of cities, which can be beneficial to related regional and policy studies.

In this paper, we first look at some basic statistics on Chinese venture capital industry, and then investigate the quantitative relationship between venture capital activity (both the number of investments and total investment amount) and the urban size (i.e., population) in China over roughly the last two decades. In particular, we identify a nonlinear scaling relation between them, which is a manifestation of emergent behavior of complex system from general micro-level interactions among the system’s constituent units \cite{west2017scale}. And in recent years, the scaling exponents for venture capital activities are larger than the ones for GMP (Gross Metropolitan Product) of cities, which suggest that venture capital activities are more complex phenomena as indicated by recent advances \cite{balland2020complex,gomez2016explaining} and have a higher concentration in larger cities.  
Furthermore, we take the scaling relationship as a baseline to make a more objective evaluation on the attractiveness to venture capitals by exploiting the Scale-Invariant Metropolitan Indicator (SIMI) that controls the nonlinear size effect induced by agglomeration and ensuing nonlinear interactions between individuals. Different from the evolution of SIMI on GMP which is quite stable over decades, the evolution of SIMI on venture capital activities are undergoing larger fluctuations, and there are three groups of cities can be identified through their evolution patterns, two of which stand out with increasing (investment-enhancing cities) and decreasing trends (investment-declining cities) over decades, respectively. The gaining or losing momentum on attracting venture capital might reflect the impacts of local policies on attracting investment which still require further detailed investigation and we assume it also manifest a process of forming order and hierarchy in a not yet mature market.  
In addition, we also notice that though scaling laws hold over time, the scaling exponents of VC activities are changing non-trivially in past decades which can be explained via different growth dynamics of cities by a conceptual 
framework proposed by us, which are important for better understanding the indications of urban scaling laws. And we assume that such a conceptual model can be general to other scenarios, including the evolution of scaling exponents on traffic congestion \cite{depersin2018global}, built-up areas, volume of urban road networks, electricity consumption \cite{xu2019paradoxical}.

\section{Results}
\subsection{Dataset and preprocessing}
We get access to detailed venture capital investment records in Chinese VC industry from SiMuTong dataset (Zero2IPO Group) \cite{website:PEdata}. The dataset we purchased is the most authoritative VC industry dataset in China. Each record in investment dataset gives us the name of investor (usually a VC firm, sometimes are an individual investor or angel), the company got invested and the company's basic information including industry and location, the date of investment, amount, share, stage, and the round of investment, etc. 

Before 2018, there were 48,700+ investment records, and we remove all records where the location of the company got invested is not in China, which corresponds to 510 records. 
For the dataset we performed some further basic processings: we query a more detailed location information (at least at the city level) from Qichacha (www.qcc.com); and convert the investment amount in foreign currency into RMB; besides, for better consistency, we also map the industry type of the start-up companies indicated by SiMuTong dataset to the one in the Industrial Classification for National Economic Activities of China (ICNEAC).

\subsection{Basic statistics of venture capital investment activities in China}
\begin{figure}[htbp]  \centering
\includegraphics[width=0.75\linewidth]{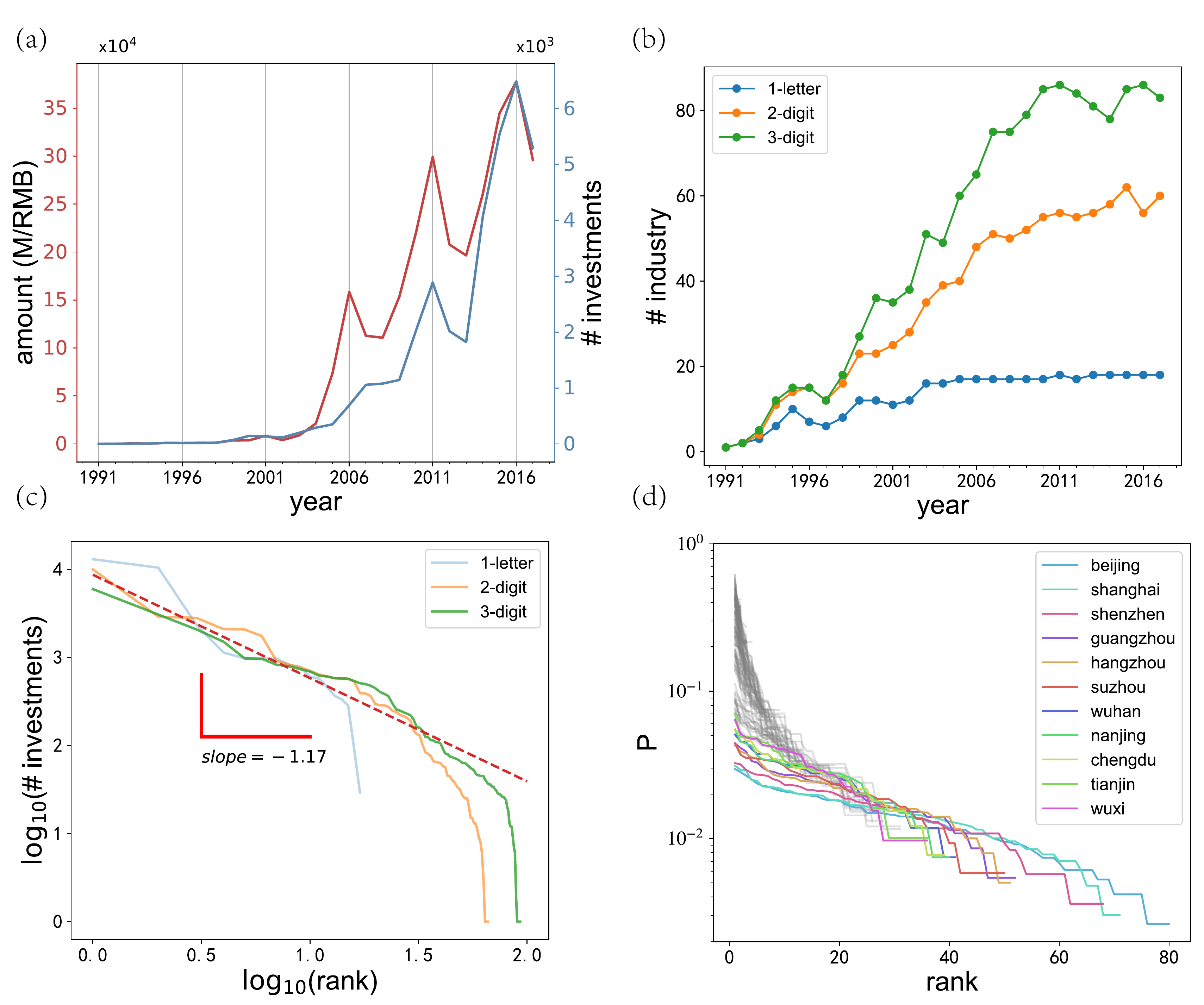}
\caption{Basic statistics of venture capital investment activities in China. (a) The size of annual investment amount (left Y axis) and the number of investment activities (right Y axis) from the year 1991 to 2017, both of which undergo fast increase after 2000, and manifest a roughly five-year cycle. (b) The number of industries got VC investment in each year at successive classification level according to the Industrial Classification for National Economic Activities of China. At first level (encoded by 1-letter), there are 17 categories; second level (2-digit) has 97 categories; third level (3-digit) has hundreds of categories. (c) The Zipfian rank distribution on the number of VC investment activities on each industry at successive classification level. $\#inv.\propto rank^{-1.17}$ at the third level, which indicates that the top ranking industry receives roughly twice of investments as of the second top ranking one, and triple of the third one, and so on. (d) The rank distribution of industries in each city, where each line represents an individual city, and the rank of each industry within a city is still calculated from the number of investments, and the Y axis corresponds to its fraction in each city.}
    \label{fig1}
\end{figure}

Since the origins of the modern private equity (PE) industry (VC is a type of PE) in 1946, there had been a first boom and bust cycle worldwide from 1982 to 1993 \cite{gompers2004venture}. While in China, VC is still a relatively newly emergent industry. Until 1985, the first Chinese VC firm (the China New-tech Venture Capital Corporation, CNVCC) was established, which was fully-owned by the Ministry of Science and Technology \cite{lu2013venture}. And the first foreign VC that entered China was IDG in 1992.  
At very first there were quite a few VC firms and investment activities.  
The size of annual investment amount and the number of investment activities undergoes fast increase after 2000, and they both manifest a roughly five-year cycle (see Fig.~\ref{fig1}(a)). The industry diversity of VC investments gradually reaches to the most of first- and second-level industrial categories (see 1-letter and 2-digit classifications in Fig.~\ref{fig1}(b)). According to the Industrial Classification for National Economic Activities of China (ICNEAC), there are 17 categories of first-level industry (encoded by 1-letter), 97 of second-level (encoded by 2-digits). There are four levels of classifications in ICNEAC, but due to the limitation on resolution and the nature of the VC investment dataset, the industry category can only be mapped to the third-level of classifications in ICNEAC (encoded by 3-digits). Furthermore, among all categories of industries got VC investments over the past three decades, we find that the number of VC investment activities in all industries follows a Zipf's law at all three successive classification levels (see Fig.~\ref{fig1}(c)) with an exponent approximately equals to -1.17, which indicates that the number of investments of top ranking industry is roughly twice (more precisely, 2.25 times) of the second top ranking industry, and roughly triple (3.62 times) of the third one, and so on (see Supplementary Figure 1 for more details of industries at the third-level classification). While within each city, especially for big cities, their rank distributions on industries are similar, where the rank is calculated from the number of investments in each industry. We can observe that though the ranking of specific industries might vary in different cities, the whole distribution and the slope between big cities are quite comparable. For smaller cities (grey lines in Fig.~\ref{fig1}(d)), the number of industries got VC investments are smaller and the slope is much steeper than in big cities, which means that the top ranking industry in these small cities has a more dominant position (see Fig.~\ref{fig1}(d)).

\subsection{Scaling law on the venture capital activities and evaluation on the attraction of venture capital}
\begin{figure}[!htbp]  \centering
\includegraphics[width=\linewidth]{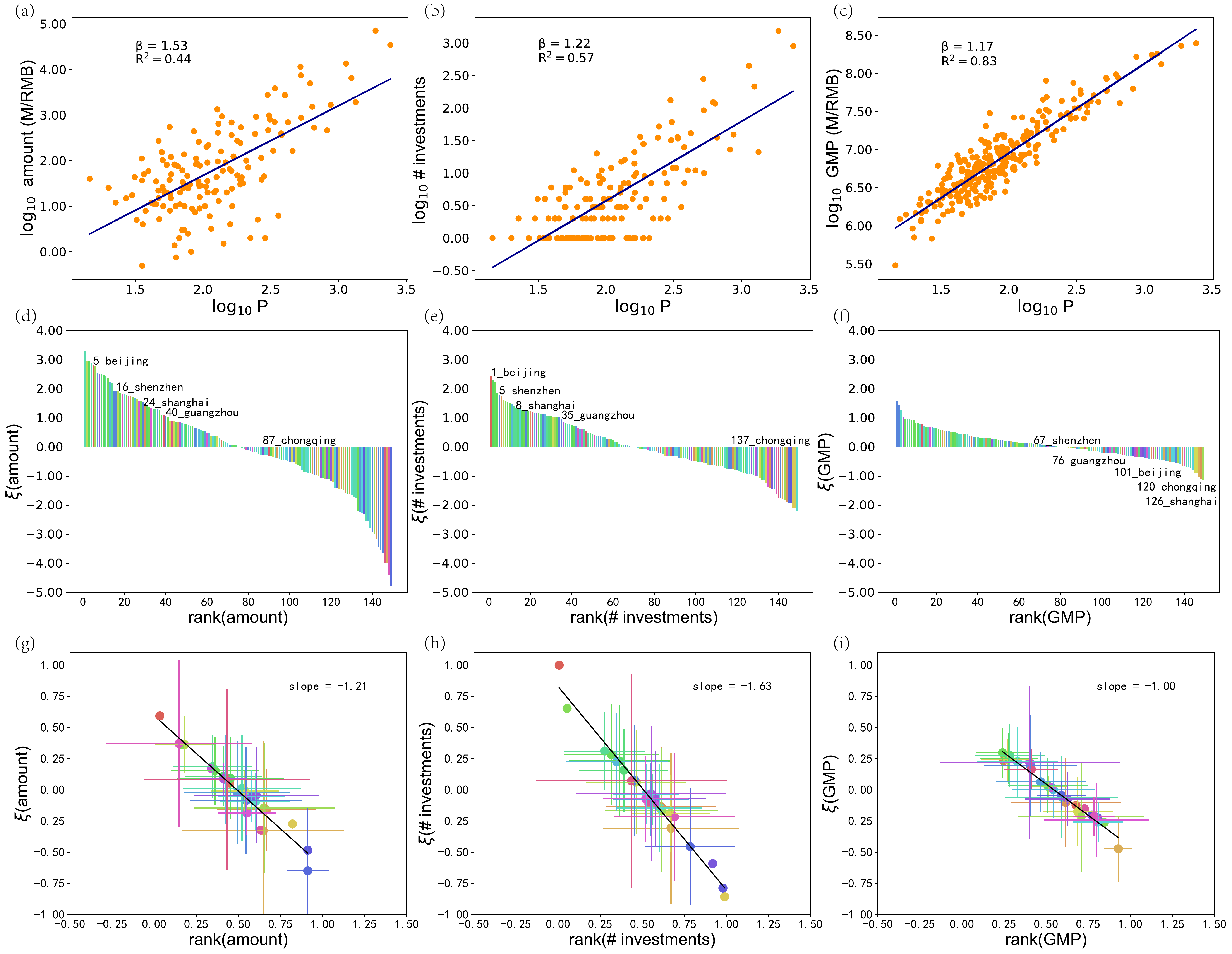} 
\caption{Urban scaling Laws on venture capital activities and the corresponding Scale-Invariant Metropolitan Indicators (SIMI). (a) The total investment amount, (b) The number of investments, (c) Gross Metropolitan Product (GMP) and population for Chinese cities in 2015. The unit of population is 10,000. Each dot represents a city, and the solid line is obtained from ordinary least squares (OLS) fitting from the data. (d-f) The corresponding SIMI plots as of (a-c). Cities are ranked according to their SIMI values. The five largest cities (measured by population) are labeled with the number indicating its rank, and cities in the same province are in the same color. (g-i) The corresponding provincial average as of (d-f). The mean and standard deviation (indicated by vertical and horizontal line attached to the node) of normalized SIMI values and ranks, where $\xi_i(t)=\xi_i(t)/\max \xi_j(t)$ and $rank_i(t)=rank_i(t)/\max rank_j(t), j\in[1,N]$. The color coding of (d-f) and (g-i) is the same. We can clearly observe that the difference between provinces are more significant on VC activities (especially on the number of investments) than on GMP, as indicated by a larger absolute magnitude of the slope.}
    \label{fig2}
\end{figure}

After a glance at the Chinese venture capital industry, we then investigate the relation between the venture capital activities and urban population, and identify a power law relation between measures of venture capital activities $Y$ and population size $P$:
\begin{equation} \label{eq.scalingLaw}
    Y_i(t)=Y_0(t)P_i(t)^\beta e^{\xi_i(t)},    
\end{equation}
where $Y_i(t)$ is the concerned urban quantity of city $i$ at time $t$ (such as the total amount of investment, the number of investments), $Y_0(t)$ is the intercept characterizing the baseline quantity per capita in the system, $P_i(t)$ is the population of city $i$, $\beta$ is the scaling exponent (or elasticity in the language of economics) which tends to be approximately independent of city size $P_i$, and $\xi_i(t)$ is the residual term. 
From Fig.~\ref{fig2}(a-c), we can observe that venture capital activity (for both of the total amount of investment and the number of investments) scales superlinearly ($\beta>1$) with respect to urban population size, which exhibits increasing returns to scale and higher concentration in large cities. 
Such scaling laws span several magnitudes, covering small cities with tens of thousand residents to mega-cities inhabited by more than tens of millions.  
And by comparing with the situation of Gross Metropolitan Product (GMP, see Fig.~\ref{fig2}(c)), the scaling exponents of VC activities are much larger (1.53 and 1.22 for the total amount and number of investments, respectively, compared to 1.17 for GMP). While a larger scaling exponent means a higher concentration in big cities, and it might also reflect that the complexity of VC activities are higher \cite{balland2020complex,gomez2016explaining}. 

As shown in Fig.~\ref{fig2}(a, b), the system is nonlinear, thus traditional per capita value is not suitable for evaluating the attraction of cities on venture capital, as the per capita value will be biased to larger cities due to the super-linearity effect (i.e., $\beta>1$). For example, let's assume two cities of population $P_1$ and $P_2$, where $P_2=2P_1$, and some concerned quantity $Y$ scales superlinearly with the population (i.e., $Y\propto P^\beta, \beta>1)$. Then the per capita value $\langle m_1\rangle$ of the smaller city would be $\langle m_1\rangle=Y_1/P_1=P_1^{\beta-1}$, as for the larger city $\langle m_2\rangle=Y_2/P_2$. As $Y_2\propto P_2^\beta$ and $P_2=2P_1$, so $\langle m_2\rangle=(2P_1)^\beta/(2P_1)=2^{\beta-1} P_1^{\beta-1}=2^{\beta-1}\langle m_1\rangle$. Since $\beta>1$, averagely speaking, $\langle m_2\rangle$ will always be larger than $\langle m_1\rangle$ due to the nonlinear effect, i.e., when the system is superlinear, the per capita value will be inherently biased towards bigger cities; while, if the system is sublinear, then it's biased towards smaller cities. This indicates that per capita value is not objective on making comparisons between entities in a nonlinear system.  

As scaling laws reported in Eq.~(\ref{eq.scalingLaw}) never appear by accidents \cite{barenblatt2003scaling}, it can be regarded as an important null/reference model where the deviation from it can be a scale invariant evaluation on the attraction of cities on venture capital. 
In Eq.~\ref{eq.scalingLaw}, $\xi_i(t)$ is the residual term (i.e., the deviation) to the expectation from the scaling relation, where  
\begin{equation} \label{eq.SIMI}
    \xi_i(t)=\ln Y_i(t)-\ln Y_0 -\beta\ln P_i(t).   
\end{equation}
The sum of residuals are expected to be zero, i.e., $\sum_{i=1}^{N}\xi_i(t)=0$ where $N$ is the number of cities in the investigated system. Other important features of residual $\xi_i$ are that they are independent of city size $P_i$ and dimensionless, which makes them a great Scale-Invariant Metropolitan Indicators \cite{zund2019growth} (previously known as ``Scale-Adjusted Metropolitan Indicators (SAMI)''\cite{bettencourt2010urban}) on evaluating the performance of cities. A positive residual ($\xi_i>0$) signifies over-performance with respect to its population size, while a negative value indicates under-performance. SIMI can separate the growth induced by increasing of population and by other local policies or features, which allow us to make direct comparison between any two cities of different size and provide meaningful rankings across the whole urban system. 
From Fig.~\ref{fig2}(d-f), we can observe that though the largest cities are under-performing or having a much lower ranking on GMP (Fig.~\ref{fig2}(f)), they are generally over-performing on venture capital activities (Fig.~\ref{fig2}(d, e)). This indicates that though their performances on GMP are not that good, big cities still attract more than expected VC investments with respect to its size. For example, Beijing is the top ranking city on the number of investments, but only ranked around 100 on GMP. 

In addition, we also look at the average situation of each province whose SIMI is obtained from averaging all its cities on both SIMI values and ranks, which are all normalized by the maximum value in the year (i.e., $\xi_i(t)=\xi_i(t)/\max\xi_j(t)$ and $rank_i(t)=rank_i(t)/\max rank_j(t), j\in[1,N]$). We can observe that the difference between provinces are more significant on VC activities (especially on the number of investments) than on GMP, as indicated by a larger absolute magnitude of the slope (see Fig.~\ref{fig2}(g-i)).

\subsection{The evolution of SIMI and classification of cities} 
\begin{figure}[!htbp]  \centering
\includegraphics[width=\linewidth]{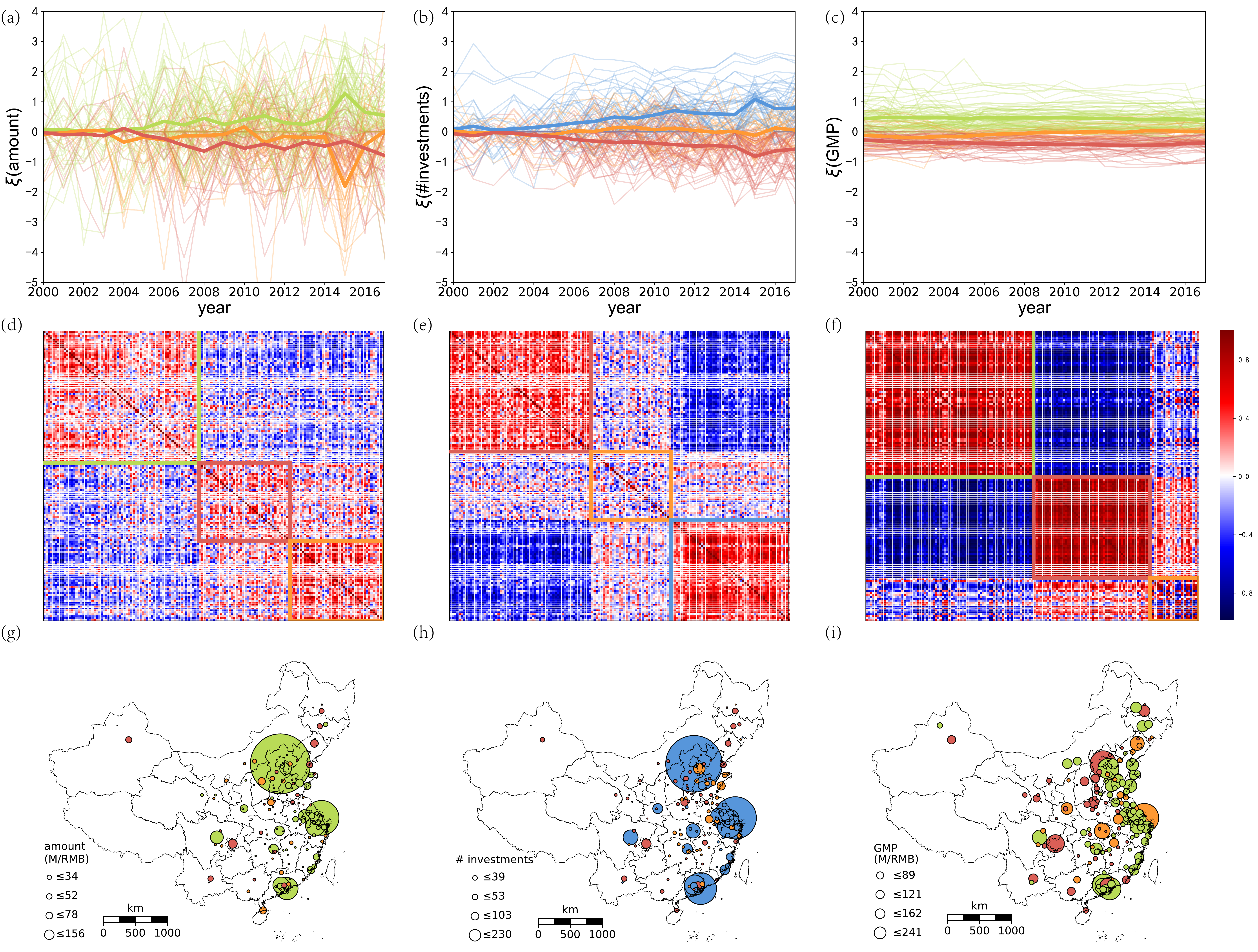}
\caption{The temporal evolution of scale-invariant metropolitan indicators (SIMIs) on (a) the total amount of investment, (b) the number of investments, (c) GMP from the year 2000 to 2017. The thicker and darker line indicates the average of the group, and each thinner and lighter line represents an individual city, and the color of the line signifies the group classification results. (d-f) The corresponding correlation matrix $C_{ij}$ calculated from the time series in (a-c). Each pixel represents the correlation $C_{ij}$ between city $i$ and $j$, i.e., each row/column stands for a city. Colored rectangle corresponds to a group obtained from standard hierarchical clustering algorithm based on the corresponding distance matrix $D_{ij}=(1-C_{ij})/2$. (g-i) The spatial distribution of cities in different groups. The size of the circle is proportional to the cumulative value of concerned quantities from the year 2000 to 2017 (i.e., the total amount of investment, the number of investments, and GMP, respectively). Note that the color coding of subfigures is only consistent in each column (e.g., subfigures (a, d, g)), not between columns.}
    \label{fig.SIMI}
\end{figure}

In addition, the temporal evolution of scale-invariant metropolitan indicators (SIMIs) on venture capital activities displays larger fluctuations than on GMP (see lighter lines in Fig.~\ref{fig.SIMI}(a-c), each of which represents a city). A larger fluctuation might indicate a fast evolving phase, while a much smoother temporal evolution also means a long-term memory. 

To infer possible connections between cities, the spatial correlation between two cities is a good indicator, which is calculated based on the cosine similarity of their SIMI time series
\begin{equation}  \label{eq.correlation}
    C_{ij}=\frac{\sum_t\xi_i(t)\xi_j(t)}{|\xi_i||\xi_j|}=\frac{\sum_t\xi_i(t)\xi_j(t)}{\sqrt{\sum_t\xi_i(t)^2}\sqrt{\sum_t\xi_j(t)^2}}.
\end{equation}
$C_{ij}$ takes off size effect and is normalized to [-1,1], where a larger positive value indicates a higher similarity (or a positive relation), and 0 means no correlation, -1 refers to the extreme case of dissimilarity (or a negative relation). Fig.~\ref{fig.SIMI}(d-f) show the correlation between cities on the time series of the amount of investment, the number of investments and GMP, respectively. 

We can further define a corresponding distance matrix $D_{ij}=(1-C_{ij})/2$, where two cities that are changing in a perfect positive trend (i.e., $\xi_i(t)=a\xi_j(t), a>0$) will be the closest with $D_{ij}=0$ (since $C_{ij}=1$); while for the two changing in a perfect negative trend (i.e., $\xi_i(t)=a\xi_j(t), a<0$), $C_{ij}=-1$ and their distance will be the longest $D_{ij}=1$. If two cities are of no correlation (i.e., $C_{ij}=0$), then the distance between them will be $D_{ij}=1/2$.  
Based on the distance matrix $D_{ij}$, we can group cities into clusters with higher similarity (shorter distance) by standard hierarchy clustering algorithm. There are three groups of cities can be identified through their evolution patterns on VC investment activities (see thicker and darker lines in Fig.~\ref{fig.SIMI}(a, b) and Fig.~\ref{fig.SIMI}(g, i) for their spatial distribution), and we can observe that all first-tier cities (i.e., Beijing, Shanghai, Guangzhou, Shenzhen) 
and the majority cities in Yangtze River Delta (comprising the areas of Shanghai, southern Jiangsu province and northern Zhejiang province) are in the same group. These cities have an increasing over-performance trend on the amount of investment (see Fig.~\ref{fig.SIMI}(d, g)) as well as on the number of investments (see Fig.~\ref{fig.SIMI}(e, h)). And the uprising groups on the amount of investment and the number of investments have a large fraction of overlapping (with a Jaccard Index equals 0.62, see Supplementary Fig. 2 for more comparison between clustering results). 
Generally, there are three groups of cities: investment-enhancing city, investment-stable city and investment-declining city, two out of which stand out with increasing (investment-enhancing cities) and decreasing trends (investment-declining cities) over decades, respectively. The gaining or losing momentum on attracting venture capital might reflect the impacts of local policies on attracting investment which still require further detailed investigation and we assume it also manifest a process of forming order and hierarchy. In comparison, there's almost no up and down of the average situation on SIMIs of GMP, where three groups are above-, on- and under-average from the prediction of scaling laws for decades, respectively. 

In addition, from Fig.~\ref{fig.SIMI}(g, h), we can clearly observe that over the past two decades, those megacities (such as Beijing, Shanghai, Guangzhou, Shenzhen) received far more investments than other cities (no matter on the total amount or the number of investments), and we find that cities in Yangtze River Delta (comprising the areas of Shanghai, southern Jiangsu province and northern Zhejiang province) and Pearl River Delta (comprised by nine cities including Shenzhen, Guangzhou and Zhuhai) are in a positive relationship and of relatively similar size (indicated by the size of the circle which represents the cumulative value from the year 2000 to 2017), however, the situation in Jing-Jin-Ji region (comprising Beijing, Tianjin and all of cities in Hebei province) is totally different, where Beijing takes an overwhelming dominant position, cities in Hebei province are of much smaller size, and have just an average-performance or even an under-performance with a declining trend on venture capital investments (see Fig.~\ref{fig.SIMI}(a, g) and Fig.~\ref{fig.SIMI}(b, h)). 
In comparison, the differences between cities on GMP in these regions are seemingly relatively smaller (see Fig.~\ref{fig.SIMI}(i)), while researchers have already gained similar impressions about cities in these regions. For example, Beijing has a strong ``siphon effects'' over cities in Hebei or even over Tianjian \cite{zhou2018addressing,hu2020roles}; while between Shanghai and other cities in Yangtze River Delta, there's a strong spatial spillover effect and coordinated development \cite{wang2020high}. Further comparative analysis might reveals deeper interacting relationship between cities on venture capital activities in these large urban agglomeration regions, and reveal more fundamental urban growth dynamics.

\subsection{The evolution of scaling exponents and its explanation}
\begin{figure}[!htbp]  \centering
\includegraphics[width=0.75\linewidth]{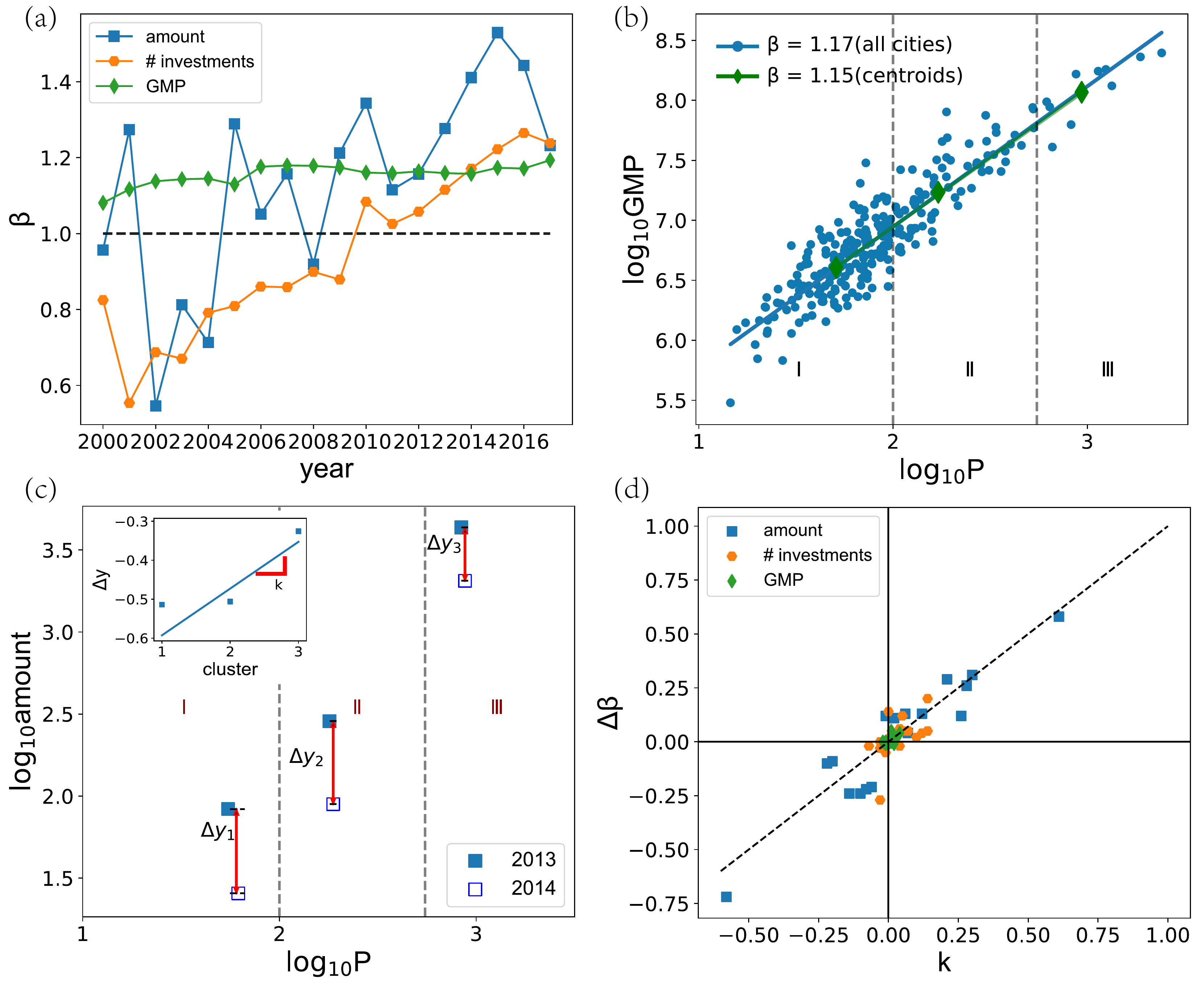}
\caption{(a) The temporal evolution of scaling exponent $\beta$ on the total amount of investment, the number of investments, GMP from the year 2000 to 2017. $R^2$ of fittings in each year are shown in Supplementary Fig. 3. (b) The scaling exponents estimated from the ensemble of the data and three centroids (denoted as green diamonds) of cities with different size, which is the average of both population and concerned variable. The estimated exponents are quite close. So, when coming to explain the evolution of scaling exponents, we exploit such centroid view. 
(c) The dynamic change measured by centroids between two consecutive years on the amount of VC investment from the year 2013 to 2014, where the population growth is relatively small and thus we focus on the difference on Y axis $\Delta y(t)$s. (inset) The fitting slope $k(t)$ of $\Delta y(t)$s of centroids for groups of cities with different size, where $\Delta y(t)=y(t)-y(t-1)$. (e) Correlation between the change of scaling exponent $\Delta \beta(t) = \beta(t)-\beta(t-1)$ and the slope $k(t)$ on the total amount of investment, the number of investments, GMP from the year 2000 to 2017, where each marker represents the result from two consecutive years. We can observe that most of the data are around the diagonal, which indicates that the growth dynamics of different sized cities can well predict the change of scaling exponent $\beta$.}
    \label{fig.ScaEvo}
\end{figure}

Apart from the larger fluctuations of SIMI values over time, we also notice a considerable change on the scaling exponents of VC activities (for both of the amount of investment and the number of investments, see Fig.~\ref{fig.ScaEvo}(a)). Compared to GMP, which is relatively stable over time, the evolution of scaling exponents $\beta$ on VC activities are of much larger fluctuations, where the overall trend is increasing and undergoing a phase change from sublinear to superlinear. Such trend also reveals that the concentration of venture capital activities in big cities is becoming stronger. 

In order to explain the dynamics leading to such evolution, we propose a conceptual model where cities are divided into three groups -- small, middle and large sized cities, and the centroids of each group are regarded as averaged representative \cite{bettencourt2020urban} (see Fig.~\ref{fig.ScaEvo}(b) where larger diamond-shape markers are the centroids, whose value on GMP and population is the average of all cities in each group, respectively). And we find that the scaling exponent estimated from the centroids is in good agreement with the one estimated from all cities (see Fig.~\ref{fig.ScaEvo}(b)). So, thereafter, we utilize such centroid representation to explain the change of scaling exponents. We assume that if smaller cities has a larger increase than bigger cities, then the scaling exponent would decrease in the next year, and vice versa.

In Fig.~\ref{fig.ScaEvo}(c), we take the situations of the year 2013 and 2014 on the amount of investment as an example, where the scaling exponent in the year 2014 is larger than the one in 2013 (see Fig.~\ref{fig.ScaEvo}(a)). 
It's worth noting that different from GMP of cities or other common socioeconomic quantities \cite{hong2020universal}, which is usually increasing (i.e., with a positive $\Delta y$ value, where $\Delta y(t) = y(t)-y(t-1)$), the situations for the total amount of investment or the number of investments can be negative, see the example in Fig.~\ref{fig.ScaEvo}(c) where the average value is smaller than previous year. In this example, although all cities are decreasing, smaller cities are dropping at a larger magnitude than bigger ones -- $|\Delta y_1(t)|\approx |\Delta y_2(t)|>|\Delta y_3(t)|$), so the scaling exponent is still increasing from the year 2013 to 2014 ($\Delta \beta(2014)=\beta(2014)-\beta(2013)>0$) which is in agreement with a positive slope $k(2014)$ regressed from $\Delta y(t)$s as shown in the inset of Fig.~\ref{fig.ScaEvo}(c). 

And more importantly, from Fig.~\ref{fig.ScaEvo}(d), we find that $k(t)$ can be quite close to the change of scaling exponent ($\Delta \beta(t)=\beta(t)-\beta(t-1)$), this further indicates that the growth dynamics of different sized cities can well explain the evolution of scaling exponents, which we assume would be able to apply to other scenarios \cite{depersin2018global,strano2016rich}.

\begin{figure}[!htbp]  \centering
\includegraphics[width=0.5\linewidth]{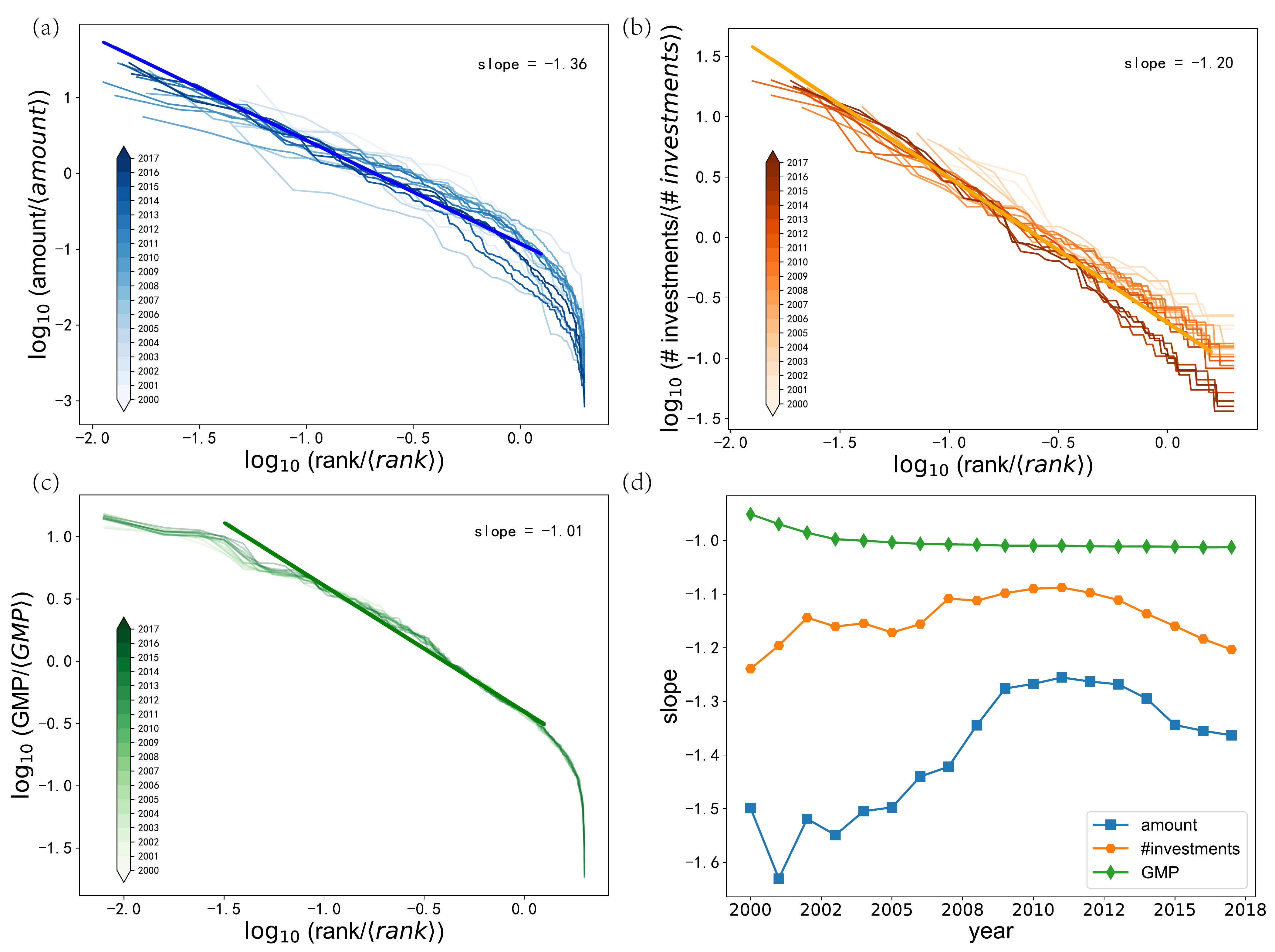}
\caption{Mean normalized Zipf's Law over past decades on (a) the total amount of investment, (b) the number of investments, (c) GMP from the year 2000 to 2017, and (d) the evolution of exponents of Zipfian distribution in (a-c). The average exponent is indicated by the line and value of the slope in (a-c). Corresponding Zipfian plots before normalization are shown in Supplementary Fig. 4.}
    \label{fig.ZipfEvo}
\end{figure}

In comparison, we can observe that fluctuations of the exponent of Zipfian distributions (see Fig.~\ref{fig.ZipfEvo}(d)) are relatively smaller than the cases on scaling law exponents reported in Fig.~\ref{fig.ScaEvo}(a). 
This further indicates that though the growth dynamics of cities can be quite different, yet the hierarchical structure of all cities is relatively stable. The relation between the evolution of Zipf's Law and scaling law requires further investigations. In addtion, we can observe that after performing a mean normalization, the Zipfian distributions are well collapsed (see Fig.~\ref{fig.ZipfEvo}(a-c)). This also indicates that such hierarchical structure is also relatively stable over time and not very sensitive to the size of the whole system.

\section{Discussion}
Though relatively new to China, venture capital industry has been coming to a booming phase with both the size and diversity increasing over past two decades. 
We first identify a significant nonlinear scaling law between VC activities and urban population. In recent years, the scaling exponents of VC activities are much larger than the exponents of GMP, which indicates that there's a stronger concentration in bigger cities, and VC activities are of higher complexity as indicated by a larger scaling exponent. 

As for evaluating the attractiveness of venture capital, in such nonlinear systems, the widely applied linear per capita indicators would be biased to larger cities if the system is superlinear, or biased to smaller cities if it is sublinear. So the SIMI of cities relative to the prediction of scaling laws is a more objective and size independent metric, which provides more meaningful ranking and is able to distinguish the effects of local dynamics and change induced by the change of population size. We find that different from GMP, where big cities are usually under-performance, big cities are usually over-performing on attracting VC investments. Besides, the evolution of SIMIs on VC activities undergoes much larger fluctuations than the case of GMP, which indicates that there's no long-term memory on attracting VC investments. 
In addition, the spatiotemporal evolution of SIMI on VC activities reveal three distinct groups of cities (investment -enhancing, -declining and -stable cities), two of which stand out with increasing and decreasing trend, respectively. And the taxonomy results also signify different development mode between large urban agglomeration regions. As we can observe that in the Jing-jin-ji region, Beijing takes an overwhelming dominant position with ``siphon effect'' over cities in Hebei province and even Tianjian; In comparison, cities in Hebei province are of much smaller size, and have just a normal performance relative to their population size on the number of investments or even under-performance with a declining trend on the amount of investment. While in Yangtze River Delta and Pearl River Delta, cities are generally in a positive relationship and of relatively similar size. Such discoveries would be informative and beneficial to related regional and policy studies. 

In addition, we observe that the scaling exponents of VC activities are changing non-trivially in past decades, with an overall increasing trend and much larger fluctuations compared to the situation of GMP. We find that such evolution can be well explained via different growth dynamics of cities by a conceptual model proposed by us. The idea behind our model is simple: if smaller sized cities have a larger increase (or smaller decrease) than bigger cities, then in the next year, the scaling exponent would be smaller, and vice versa. We find that with the simple measure developed by us, the change of scaling exponents can be reasonably well predicted by our model, which focus on the growth dynamics of different sized cities. We assume that our model would be general to other scenarios, including the evolution of scaling exponents on traffic congestion, built-up areas, volume of urban road networks, electricity consummations. In comparison, the evolution of Zipfian distributions are smoother, which indicates that the hierarchical structure between cities are relatively stable over time.  

An important future work would be integrating identified scaling relationships into a predictive theory of endogenous (population and economic) growth model of cities. The relationship between venture capital activities and patenting in cities can be another important future work. 



\subsection*{Data accesibility.} All data needed for reproducing our analysis are available at \url{https://github.com/UrbanNet-Lab/UrbanScaling_VentureCapital/tree/master/data}. The raw data on venture capital investments is purchased from SiMuTong dataset of Zero2IPO Group (\url{www.pedata.cn}), for which we cannot disclose. 
Complemented information of start-up companies can be queried from Qichacha (\url{www.qcc.com}). The code for reproducing our analysis can be found at \url{https://github.com/UrbanNet-Lab/UrbanScaling_VentureCapital/tree/master/code}. 

\subsection*{Authors' contributions.}
R.L. and H.E.S. conceived the study, L.L., G.X. and S.M. cleaned and complemented the data,  L.L. performed analysis, R.L. and L.L. analysed the results. R.L. was the lead writer of the manuscript, and all authors reviewed the manuscript.

\subsection*{Competing interests.} The authors declare no conflict of interests. 


\subsection*{Acknowledgements.}
We acknowledge financial supports from the National Natural Science Foundation of China (Grant No. 61903020), BUCT Talent Start-up Fund (Grant No. BUCTRC201825). 
H.E.S. acknowledges financial support from NSF Grant PHY-1505000 and Defense Threat Reduction Agency Grant HDTRA1-14-1-0017. L.L. acknowledges helpful discussion with Tianyu Cui at UrbanNet Lab.



\end{document}